\documentclass[10pt]{article}
\usepackage{OSAmeet}
\usepackage{graphicx}

\begin{document}

\title{Analysis of the ellipticity induced PMD and general scaling perturbation in a transmitting fiber.}

\author{Maksim Skorobogatiy, Mihai Ibanescu, Steven Johnson, Ori Weisberg, Torkel Engeness, Steven Jacobs, Yoel Fink}
\address{OmniGuide Communications, One Kendall Square, Build.
100, Cambridge, MA 02139, USA.} \email{maksim@omni-guide.com}


\begin{abstract}
Presented is an analysis of general scaling perturbations in a
transmitting fiber. For elliptical perturbations, under some
conditions an intermode dispersion parameter characterizing modal
PMD is shown to be directly proportional to the mode dispersion.
\end{abstract}


\noindent                      

In the following paper we derive a generalized Hermitian
Hamiltonian approach for the treatment of Maxwell equations in
waveguides as well as develop a perturbation theory for the
general class of scaling perturbations that include ellipticity
and a uniform scaling of an arbitrary index profile. Because of
the Hermitian nature of the formulation most of the results from
the well developed perturbation theory of quantum mechanical
systems can be directly related to the light propagation in the
waveguides. Such formulation provides for an intuitive way of
understanding PMD and birefringence in the elliptically perturbed
fiber profiles. Region of validity of our theory extends to the
case of large variations of the dielectric constant across the
fiber crossection and is limited only by an amount of re-scaling.
Finally, we establish that if in some range of frequencies a
particular mode behaves like a mode of pure polarization $TE$,$TM$
(where polarization is judged by the relative amounts of the
electric and magnetic longitudinal energies in a modal
crossection) its inter-mode dispersion parameter
$\tau=|\frac{\partial\bigtriangleup\beta_{e}}{\partial\omega}|$ is
related to its dispersion $D$ as $\tau=\lambda\delta |D|$, where
$\delta$ is a measure of the fiber ellipticity and
$\bigtriangleup\beta_{e}$ is a split in a wavevector of a linearly
polarized doubly degenerate mode of interest due to an elliptical
perturbation.

While there has been a wide amount of work done on estimating such
quantities as local birefringence induced by perturbations in the
fiber profile most of the treatments were geared toward
understanding the low contrast, weakly guiding systems such as
ubiquitous silica waveguides and are not directly applicable to
the high contrast systems such as Bragg fibers, photonic crystal
fibers and integrated optics waveguides which are steadily
emerging as an integral part of the state of the art transmission
systems.

In deriving Hamiltonian formulation for the eigen fields of a
generic waveguide exhibiting translational symmetry in
longitudinal $\hat{{\bf z}}$ direction we start with a well known
set of Maxwell equations written in terms of transverse and
longitudinal fields \cite{jackson}. Assuming the form of the field
\begin{equation}
\left |
\begin{array}{c}
{\bf E}(x,y,z,t)\\
{\bf H}(x,y,z,t)
\end{array}
\right | = \left |
\begin{array}{c}
{\bf E}(x,y)\\
{\bf H}(x,y)
\end{array}
\right |exp(i\beta z-i\omega t) \label{eq:symmetry}
\end{equation}
and introducing the transverse and longitudinal components of the
fields as ${\bf F}={\bf F}_t+{\bf F}_z$, ${\bf F}_z={\bf \hat{z}}
F_z$, ${\bf F}_t=({\bf \hat{z}}\times {\bf F}) \times {\bf
\hat{z}}$ Maxwell equations can be rewritten as a generalized
Hermitian eigen problem \cite{johnson}
\begin{equation}
\beta \left \|
\begin{array}{cc}
0 & -{\bf \hat{z}}\times\\
{\bf \hat{z}}\times & 0\\
\end{array}
\right \| \left |
\begin{array}{c}
{\bf E}_t\\
{\bf H}_t\\
\end{array}
\right |= \left \|
\begin{array}{cc}
\omega\epsilon - \frac{1}{\omega}{\bf \bigtriangledown}_t\times
[{\bf \hat{z}}({\bf \hat{z}}\cdot({\bf \bigtriangledown}_t \times))] & 0\\
0 & \omega - \frac{1}{\omega}{\bf \bigtriangledown}_t\times
[{\bf\hat{z}}(\frac{1}{\epsilon}{\bf\hat{z}}\cdot({\bf
\bigtriangledown}_t \times))]
\end{array}
\right \| \left |
\begin{array}{c}
{\bf E}_t\\
{\bf H}_t\\
\end{array}
\right |. \label{eq:Maxwell}
\end{equation}
In this form operators on the left and on the right are Hermitian
thus defining a generalized Hermitian eigen problem and allowing
for all the convenient properties pertaining to such a form,
including real eigenvalues $\beta$ as well as orthogonality of the
modes corresponding to the different $\beta$'s (for more
discussion see \cite{johnson}). Defining hermitian operator on the
left of (\ref{eq:Maxwell}) as $\hat{B}$ and the one on the right
$\hat{A}$ and introducing Dirac notation $|\psi> = \left |
\begin{array}{c}
{\bf E}_t\\
{\bf H}_t\\
\end{array}
\right |$ we rewrite a generalized eigen problem as
\begin{equation}
\beta \hat{B}|\psi_{\beta}>=\hat{A}|\psi_{\beta}>
\end{equation}
with a condition of orthogonality between modes of $\beta$ and
$\beta'$ in the form
\begin{equation}
<\psi_{\beta}|\hat{B}|\psi_{\beta'}>=\delta_{\beta,\beta'}.
\label{eq:orthogonality}
\end{equation}

In the following we analyze uniform along ${\bf{\hat{z}}}$ axis
perturbations. If a perturbation such as general re-scaling of
coordinates $x_{scaled}=x(1+\delta_x)$,$y_{scaled}=y(1+\delta_y)$
is introduced into the system it will modify an operator
$\hat{A}$. A particular case of general re-scaling when
$\delta_x=\delta_y$ correspond to the uniform scaling of a
structure, while the case of  $\delta_x=-\delta_y$ corresponds to
a uniform ellipticity. Denoting a correction to an original
operator on the left of (\ref{eq:Maxwell}) $\delta\hat{A}$, the
new eigen values $\tilde{\beta}$ of the split doubly degenerate
eigen mode are found by solving a secular equation \cite{landau}
and gives
\begin{equation}
\beta^{\pm}=\beta+\frac{<\psi_{\beta,m}|\delta\hat{A}|\psi_{\beta,m}>}
{<\psi_{\beta,m}|\hat{B}|\psi_{\beta,m}>}
\pm\frac{<\psi_{\beta,m}|\delta\hat{A}|\psi_{\beta,-m}>}{|<\psi_{\beta,m}|\hat{B}|\psi_{\beta,m}>|}.
\label{eq:pert_deg}
\end{equation}
The inter-mode dispersion parameter being proportional to PMD
\cite{PMD} is defined to be equal to the mismatch of the inverse
group velocities of the split due to the perturbation modes
$\tau=\frac{1}{v^+_g}-\frac{1}{v^-_g}$ which is, in turn, can be
expressed in terms of the frequency derivative
$\tau=\frac{\partial(\beta^+-\beta^-)}{\partial\omega}$.

Now we derive a form of the perturbation operator for the cases of
uniform scaling and uniform ellipticity. We start with an
elliptical waveguide and a generalized Hermitian formulation
(\ref{eq:Maxwell}) where the derivatives in operator $\hat{A}$
should be understood as the derivatives over the coordinates
$x_{scaled}$ and $y_{scaled}$. We then transform into the
coordinate system in which an elliptical waveguide becomes
cylindrical. Assuming normalization (\ref{eq:orthogonality}) after
some combersome algebra \cite{skorobogatiy} we arrive at the
following expressions.

Case of uniform scaling $\delta_x=\delta_y=\delta$
\begin{equation}
\begin{array}{l}
\bigtriangleup\beta_s=<\psi_{\beta,m}|\delta\hat{A}|\psi_{\beta,m}>=
2\delta\int_{S}ds \left |
\begin{array}{c}
E_r\\
E_\theta\\
H_r\\
H_\theta\\
\end{array}
\right |^+_{\beta,m} \left \|
\begin{array}{cccc}
\omega\epsilon & 0 & 0 & -\beta\\
0 & \omega\epsilon & \beta & 0\\
0 & \beta & \omega & 0\\
-\beta & 0 & 0 & \omega\\
\end{array}
\right \| \left |
\begin{array}{c}
E_r\\
E_\theta\\
H_r\\
H_\theta\\
\end{array}
\right |_{\beta,m}=2\eta\omega\int_{S}ds (\epsilon|E_z|^2+|H_z|^2)
\end{array}
\label{eq:db_scaling}
\end{equation}

Another important result about change in the propagation constant
of a mode under uniform scaling is that its frequency derivative
is proportional to the dispersion of a mode. To derive this result
we consider a dispersion relation for some mode of a waveguide
$\beta=f(\omega)$. From the form of Maxwell equations it is clear
that if we uniformly re-scale all the transverse dimensions in a
system by a factor of $(1+\delta)$ then the new
$\tilde{\beta}=\beta+\bigtriangleup\beta_s$ for the same $\omega$
will satisfy $\tilde{\beta}=\frac{f(\omega(1+\delta))}{1+\delta}$.
Decomposing the last expression in Taylor series and collecting
terms of the same order in $\delta$ we derive expressions for
$\bigtriangleup\beta_s$ and its derivative $\bigtriangleup\beta_s=
\delta(\omega\frac{\partial\beta}{\partial\omega}-\beta)$,
$\frac{\partial \bigtriangleup\beta_s}{\partial
\omega}=\delta\omega\frac{\partial^2 \beta}{\partial
\omega^2}=-\lambda\delta D(\omega)$ where $D(\omega)$ is a
dispersion of the mode.

Case of uniform ellipticity $\delta_x=-\delta_y=\delta$. A first
order correction to the split in the values of propagation
constants of the modes $(\beta,1)$ and $(\beta,-1)$ due to the
uniform re-scaling becomes (\ref{eq:pert_deg})
\begin{equation}
\begin{array}{l}
\bigtriangleup\beta_e=2|<\psi_{\beta,1}|\delta\hat{A}|\psi_{\beta,-1}>|=
2\eta\omega|\int_{S}ds \left |
\begin{array}{c}
E_r\\
E_\theta\\
H_r\\
H_\theta\\
\end{array}
\right |^+_{\beta,1} \left \|
\begin{array}{cccc}
-\epsilon & -i\epsilon & 0 & 0\\
-i\epsilon & \epsilon & 0 & 0\\
0 & 0 & -1 & -i\\
0 & 0 & -i & 1\\
\end{array}
\right \| \left |
\begin{array}{c}
E_r\\
E_\theta\\
H_r\\
H_\theta\\
\end{array}
\right |_{\beta,-1}|=\\
2\eta\omega|\int_{S}ds[(-\epsilon|E_z|^2+|H_z|^2)+2Im(\epsilon
E^*_r E_\theta-H^*_r H_\theta)]|
\end{array}
\label{eq:db_ellipticity}
\end{equation}
where $E$'s and $H$'s are those of the $(\beta,1)$ mode.

From expression (\ref{eq:db_ellipticity}) we find that the split
between the degenerate modes due to the ellipticity is
proportional to the difference in the longitudinal magnetic and
electric energies in the crossection of a fiber. The rest of the
crossterms in expression (\ref{eq:db_ellipticity}) usually do not
contribute substantially to the split, unless special structures
are considered where longitudinal magnetic and longitudinal
electric energies are of the same order.

An important conclusion about PMD of a structure can be drawn when
electric or magnetic longitudinal energy dominates substantially
over the other (for a longer discussion see \cite{skorobogatiy}).
In the case of pure-like $TE$ ($E_z\sim0$) or $TM$ ($H_z\sim0$)
modes split due to the uniform scaling (\ref{eq:db_scaling})
becomes almost identical to the split in the degeneracy of the
modes due to the uniform ellipticity perturbation
(\ref{eq:db_ellipticity}). Thus, in the case when the mode is
predominantly $TE$ or $TM$ as judged by the amounts of the
corresponding longitudinal energies in the crossection we expect
$\bigtriangleup\beta_s=\bigtriangleup\beta_e$. As PMD is
proportional to the intermode dispersion parameter
$\tau=\frac{\partial\bigtriangleup\beta_e}{\partial \omega}$ and
taking into account expressions for the frequency derivatives of
$\bigtriangleup\beta_s$ we arrive to the conclusion that for such
a mode PMD is proportional to the despersion of a mode
\begin{equation}
\tau=|\frac{\partial\bigtriangleup\beta_e}{\partial
\omega}|=|\frac{\partial\bigtriangleup\beta_s}{\partial
\omega}|=\lambda\delta |D(\omega)| \label{eq:ellip_disp}
\end{equation}

\begin{figure}
\centerline{\scalebox{.5}{\includegraphics{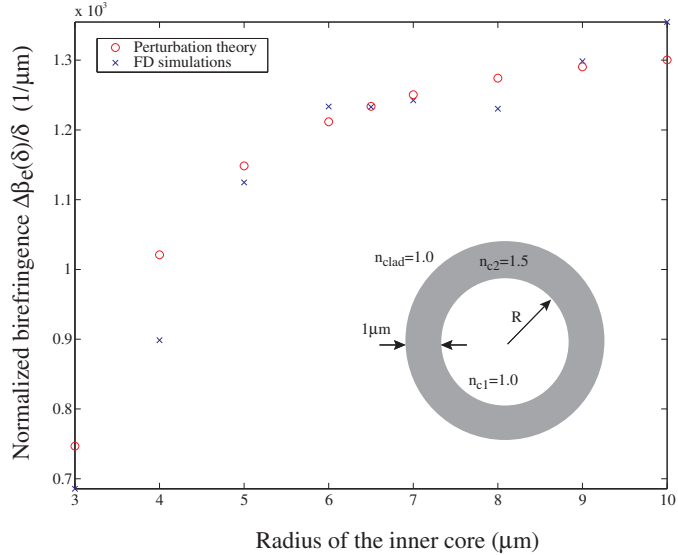}}}
\caption{\label{fig:fig_ring} Effective birefringence due to the
elliptical perturbation in a double core high dielectric contrast
fiber. Data is presented for the fundamental linearly polarized
doubly degenerate mode. While the width of a ring was kept at a
constant $1\mu m$ corresponding inner radii $R$ was varied in the
interval (3,10)$\mu m$. Split in a wavevector of an originally
degerate mode due to the uniform elliptical perturbation of
magnitude $\delta$ as predicted by the perturbation theory
(circles) is compared to the results of the Finite Difference
numerical simulations (crosses). Excellent correspondence over the
whole range of inner radii is observed.}
\end{figure}

We conclude by presenting the results of calculations of the
normalized birefringence due to the uniform elliptical
perturbation in the case of a double core high dielectric contrast
fiber Fig.\ref{fig:fig_ring}. Fundamental doubly degenerate mode
of $m=1$ was studied in a ring-like fiber of $1.0$ to $1.5$ index
contrast. Excellent correspondence between the predictions of the
pertubation theory and, in principle, exact Finite Difference
numerical simulations is observed.

\end{document}